\documentclass[letterpaper]{article} 
\usepackage{aaai23}  
\usepackage{times}  
\usepackage{helvet}  
\usepackage{courier}  
\usepackage[hyphens]{url}  
\usepackage{graphicx} 
\usepackage{natbib}  
\usepackage{caption} 
\usepackage{algorithm}
\usepackage{algorithmic}

\usepackage{newfloat}
\usepackage{listings}
\DeclareCaptionStyle{ruled}{labelfont=normalfont,labelsep=colon,strut=off} 
\floatstyle{ruled}
\newfloat{listing}{tb}{lst}{}
\floatname{listing}{Listing}

\usepackage{tabularx}
\usepackage{booktabs}

\graphicspath{ {figures/} }
\title{aedFaCT: Scientific Fact-Checking Made Easier\linebreak via Semi-Automatic Discovery of Relevant Expert Opinions}
\author{
    Enes Altuncu\textsuperscript{\rm 1}\footnote{Corresponding co-authors: Enes Altuncu (ea483@kent.ac.uk) and Shujun Li (S.J.Li@kent.ac.uk).},
    Jason R.C. Nurse\textsuperscript{\rm 1},
    Meryem Bagriacik\textsuperscript{\rm 2},
    Sophie Kaleba\textsuperscript{\rm 2},
    Haiyue Yuan\textsuperscript{\rm 1},
    Lisa Bonheme\textsuperscript{\rm 2},
    Shujun Li\textsuperscript{\rm 1}\footnotemark[1]
}
\affiliations{
    \textsuperscript{\rm 1}Institute of Cyber Security for Society (ICSS) \& School of Computing, University of Kent\\
    \textsuperscript{\rm 2}School of Computing, University of Kent\\

}

\begin{document}

\maketitle

\begin{abstract}
In this highly digitised world, fake news is a challenging problem that can cause serious harm to society. Considering how fast fake news can spread, automated methods, tools and services for assisting users to do fact-checking (i.e., fake news detection) become necessary and helpful, for both professionals, such as journalists and researchers, and the general public such as news readers. Experts, especially researchers, play an essential role in informing people about truth and facts, which makes them a good proxy for non-experts to detect fake news by checking relevant expert opinions and comments. Therefore, in this paper, we present \emph{aedFaCT}, a web browser extension that can help professionals and news readers perform fact-checking via the automatic discovery of expert opinions relevant to the news of concern via shared keywords. Our initial evaluation with three independent testers (who did not participate in the development of the extension) indicated that aedFaCT can provide a faster experience to its users compared with traditional fact-checking practices based on manual online searches, without degrading the quality of retrieved evidence for fact-checking. The source code of aedFaCT is publicly available at \url{https://github.com/altuncu/aedFaCT}.
\end{abstract}

\section{Introduction}

The digital age has evolved into an infodemic age, with the rapid propagation of false and misleading information in this highly digitised world, mixed with true and reliable information. As part of this, fake news prevents society from obtaining accurate information based on real evidence. The COVID-19 pandemic has demonstrated how fake news can seriously cause harm to people~\cite{bbc2020hund}.

Considering the amount of information available online and how fast information can be widely disseminated with the help of digital communication technologies, detecting fake news at scale is an important task. Therefore, many researchers have studied automated fact-checking methods~\cite{zhou2020}. However, automated fact-checking solutions are yet to be sufficient to adapt to various contexts, languages, and modalities. In addition, they insufficiently consider human factors, such as trust and usability, which is crucial for practical use~\cite{das2023}. These paved the way for semi-automated solutions that attempt to combine human and machine intelligence. With this respect, the literature involves many fact-checking systems leveraging human-machine teaming in different ways~\cite{guo2020, das2023}. Besides, there exists a wide range of tools and services that can assist professionals and common readers with fake news detection~\cite{nakov2021}.

Experts play a crucial role in the fight against fake scientific news by enlightening society with truths and facts through various communication channels, especially the news media. In the complicated landscape of the infodemic age, people are likely to seek help from experts they trust, such as scientists and professionals, since they are often considered the highly trusted groups in society~\cite{ipsos2022ipsos}. This makes them a proxy for non-experts to fact-check suspicious scientific claims. Other than giving expert comments and being interviewed, experts support journalists in reporting online false information to bridge gaps in their contextual understanding and methodological expertise~\cite{mcclure2022}. Besides, experts are a crucial element of the fact-checking process conducted by human fact-checkers~\cite{graves2017}. With this respect, there has been much effort to engage with experts for scientific fact-checking. To exemplify, Science Media Centre\footnote{\url{https://www.sciencemediacentre.org/}} aims to build bridges between experts and journalists so that scientific information covered in the media becomes accurate and evidence-based. Another example is Meedan's Digital Health Lab\footnote{\url{https://meedan.com/programs/digital-health-lab}}, which is composed of scientists, content moderation experts, and journalists to support evidence-based responses to health misinformation.

However, this expert-journalist collaboration could be insufficient to combat fake scientific news due to several reasons, including challenges in scientific communication~\cite{bucchi2017credibility}, the existence of outlier experts who do not share the majority opinions, and the selection of experts with incompatible expertise~\cite{palmer2020scientific}. These problems indicate the need and potential usefulness of tools that can leverage multiple experts' opinions as evidence for fact-checking purposes. Therefore, in this work, we present \emph{aedFaCT}, a web browser extension that can help professionals and common readers to discover the opinions of multiple experts on relevant topics of a particular scientific news article in a semi-automated manner. aedFaCT extracts expert opinions from several credible news sources based on a number of candidate keywords automatically extracted from the target news article, and it also automatically retrieves relevant peer-reviewed scientific publications based on such keywords. Based on the results, users can make a decision on the veracity of suspicious claims on their own by considering the retrieved evidence. Moreover, aedFaCT enables users to see a list of researchers with relevant expertise based on their publications in order to inform them about who to follow and to approach on a specific topic. In a nutshell, aedFaCT is a ``smart search assistant'' for fact-checkers to help minimise the manual work they have to do using online search engines and other known information sources.

The rest of the paper is organised as follows. Section~\ref{sec:related-work} briefly reviews related work. Then, Section~\ref{sec:mental-process} presents a focus group study to understand the mental process of users during scientific fact-checking. The architecture of the proposed system is introduced in Section~\ref{sec:system-design}, and the details of its evaluation are provided in Section~\ref{sec:evaluation}. Finally, the paper is concluded with a brief discussion in Section~\ref{sec:discussions} and the concluding remarks in Section~\ref{sec:conclusion}.

\section{Related Work}
\label{sec:related-work}

\subsection{Human-Machine Teaming Approaches in Fact-Checking}

Automated fact-checking at scale is a challenging task. Hence, recent research includes hybrid solutions based on human-machine teaming to assist fact-checkers and the general public with a level of automation in the process of fact-checking. For example, \citet{nguyen2018} designed a mixed-initiative approach to fact-checking where the system predicts the veracity of a claim based on relevant articles with their stance towards the veracity of the claim and the reputation of each source. The users' role in this design is to change the source reputation and stance of each article for more accurate prediction. More recently, \citet{gupta2021} introduced an evidence retrieval approach to search for semantically-similar news articles to assist users when validating news articles. This system leaves the fact-checking decision to the user. Moreover, \citet{labarbera2022} proposed a hybrid human-in-the-loop framework for the veracity assessment of claims, relying on three major components: AI, crowdsourcing, and experts. The veracity of the claim is considered as correctly classified if any component produces a prediction with a high confidence score. Otherwise, the claim is forwarded to the next component. Another human-in-the-loop AI system is HAMLET, a conceptual framework leveraging AI-expert teaming in multiple fact-checking tasks, such as the collection of expert data annotations and expert feedback, AI system performance monitoring, and life cycle management~\citep{logically2022hamlet}. Finally, \citet{david2021} introduced a toolbox, namely Ms.\@W, combining several publicly available services and tools that help users with fact-checking and source credibility assessment.

As another way of human-machine teaming, several fact-checking systems utilise crowd intelligence in different stages. For instance, \citet{vo2018} leveraged guardians, who are social media users correcting false information by referring to fact-checking URLs, and presented a fact-checking URL recommendation model to motivate them to engage more in fact-checking activities. Furthermore, social media companies enable users to flag posts containing false information and sent them to fact-checkers for further investigation if there are sufficient flags. Recently, Twitter launched Community Notes\footnote{\url{https://help.twitter.com/en/using-twitter/community-notes}} (previously known as Birdwatch), where users can add context to tweets to prevent the platform from false information.

\subsection{Web Browser Extensions for Fact-Checking}

Web browser extensions are quite useful for fact-checking, especially for web-based documents and articles. For example, \emph{BRENDA} allows users to perform automatically fact-checking a news article or a snippet from the article~\cite{botnevik2020}. It identifies the check-worthy claims, classifies them with a deep neural network, and then, shows the results to the user along with the evidence found from top-10 Google Search results. Another automated solution is \emph{FADE}, which discovers multiple sources containing the same news story and performs automated fact-checking according to the trustworthiness of the news sources and the cited sources in the article~\cite{jabiyev2021}. Other than the solutions developed in academia, The Factual\footnote{\url{https://www.thefactual.com/}} automatically rates news articles based on several characteristics, including their source quality and bias, author expertise, and tone.

There also exist Web browser extensions helping users with content analysis and evidence retrieval for fact-checking. One such tool is \emph{InVID}, which helps users verify videos and images with a number of tools it contains~\cite{teyssou2019}. As another example, \emph{News2PubMed} retrieves relevant health research papers given a news article~\cite{wang2021}. Another tool is called \emph{News Scan}, which shows several characteristics of the source and content of news articles, such as source popularity, sentiment, objectivity, and bias, to assist users to make a judgement on the source and content credibility~\cite{kevin-etal-2018-information}. Finally, NewsGuard\footnote{\url{https://www.newsguardtech.com/}} shows manually assigned source credibility ratings next to links on search engines and social media platforms.

\section{Mental Process of Users During Fact-Checking}
\label{sec:mental-process}

In this study, our aim is to develop a semi-automated fact-checking system for both professionals and common readers, which automates, at least, part of the users' claim investigation process. To this end, we need to understand how users manually perform fact-checking and what strategies they normally use to investigate a claim. From a general perspective, content is the most important factor for users during fact-checking~\cite{pidikiti2020}. Users mainly rely on their own knowledge and sense of judgement to make a decision, and they perform external acts of authentication (e.g., searching for more information via Google, family and friends, and experts) only if the first phase fails~\cite{tandoc2018, freiling2019detecting}. When users seek external information, they commonly prefer information that they consider credible, such as peer-reviewed scientific papers, fact-checking reports, mainstream news articles, and Wikipedia entries~\cite{he2022}.

Since the current literature lacks a systematical discussion of how different processes that fact-checkers and common readers follow to verify scientific information, we conducted a focus group discussion between the first author and three other co-authors (the third, fourth and sixth) of this paper, who were all PhD students in Computer Science focusing on a relevant research topic (AI, NLP, and/or cyber security), to understand how users verify the veracity of news content. At the time of the discussion, only the first co-author knew about the details of the study as the initialiser of the work. During the discussion, an example news article containing a false claim about COVID-19 was provided to the participants, and the investigation of the claim has been performed by discussing each step of the fact-checking process. The discussion was conducted with three fact-checking scenarios, separately: (1) the participants (as researchers) performed fact-checking themselves; (2) the participants simulated how common readers with less domain knowledge would perform fact-checking without using expert opinions as a proxy; and (3) the participants simulated how common readers would perform fact-checking by using expert opinions as a proxy. For all the scenarios, the discussion was made with the same participants instead of separate groups of researcher and common reader participants, for the sake of simplicity and to allow cross-scenario alignment. Using researchers as common readers is not necessarily a problematic setup, since researchers are effectively like common readers for research areas beyond their own expertise (e.g., health and medicine for all the authors of this paper).

In the first scenario, the participants suggested identifying some keywords about the investigated claim and using them to search for relevant research papers on Google Scholar. Then, they suggested reading the abstracts of the first few publications to make a decision, provided that they trust the publisher. In the second scenario, however, they preferred to use Google Search to search for relevant material with the same set of keywords, assuming that common readers would have been unfamiliar with scientific papers and research databases. Then, they wanted to check out the search results that are trustable for them, e.g., a news article from a news outlet they trusted, or a post from a university's official website advertising their research. Finally, in the third scenario, the participants suggested identifying multiple relevant domain experts through the websites of the corresponding institution or departments of well-known universities. Moreover, they found relevant news articles useful to identify some domain experts by checking who has been interviewed in the article.

The focus group discussion provided three major conclusions on users' scientific fact-checking process, supporting the findings of existing literature on the general fact-checking practices of fact-checkers and laypeople~\cite{juneja2022, micallef2022, he2022}: (1) domain experts were generally at the core of the fact-checking process, either explicitly, or implicitly through their publications; (2) only the sources they trusted were considered; and (3) multiple sources were taken into account for cross-checking what has been obtained.

\section{System Design}
\label{sec:system-design}

\subsection{Overview}

\begin{figure*}[tb]
\centering
\includegraphics[width=0.8\linewidth]{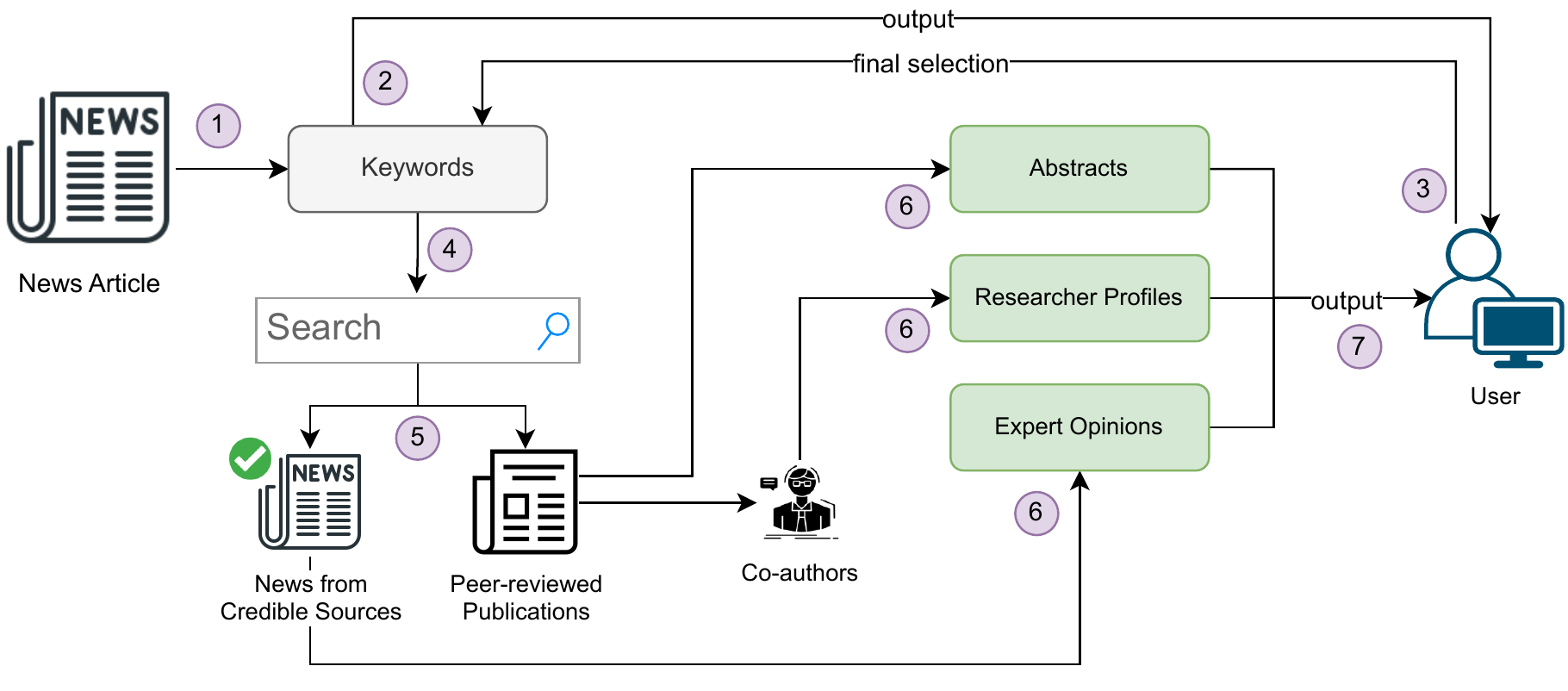}
\caption{The architecture of aedFaCT}
\label{fig:overview}
\end{figure*}

The overview architecture of aedFaCT is shown in Figure~\ref{fig:overview}. The system involves three main parts: (i) keyword extraction and selection; (ii) expert opinion discovery; (iii) scientific evidence retrieval.

\subsection{Keyword Extraction and Selection}

\begin{figure*}[tb]
\centering
\includegraphics[width=0.8\linewidth]{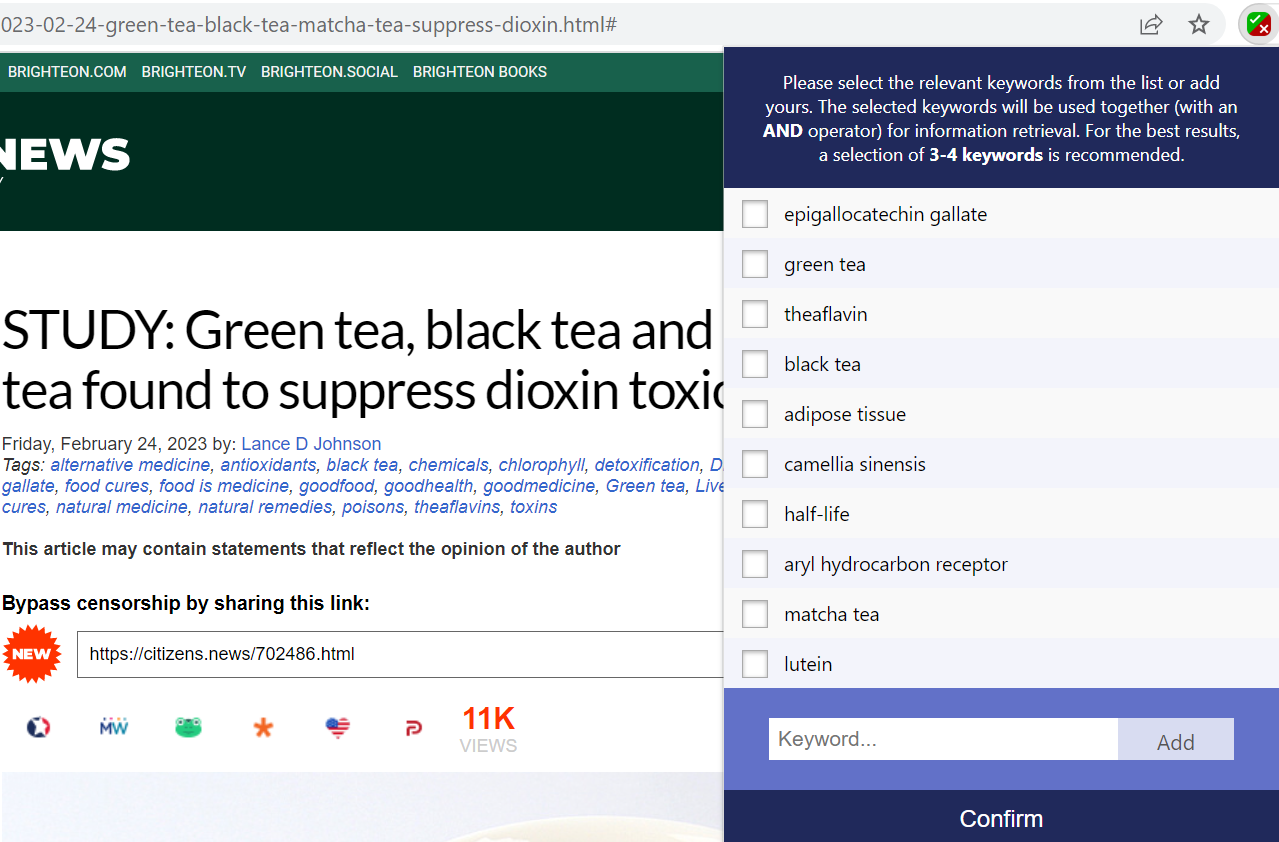}
\caption{The user interface of the keyword extraction step in aedFaCT}
\label{fig:keywords}
\end{figure*}

As the first step, the system needs to learn the context of the given news article by extracting a number of descriptive keywords. We designed this process as a human-in-the-loop mechanism to avoid topic drift while using the obtained keywords in searching. The system first fetches and parses the news content using the Newspaper3k\footnote{\url{https://newspaper.readthedocs.io/en/latest/}} library. Then, it performs automatic keyword extraction (AKE) with a state-of-the-art AKE algorithm, SIFRank+~\cite{sun2020}, to obtain the initial set of keywords. Based on the findings of our previous study~\cite{altuncu2022improving}, we used our own version of SIFRank+, enhanced with post-processing. More precisely, the enhancement involves PoS-tagging-based filtering, and prioritising keywords contained in the corresponding domain thesaurus or Wikipedia as an entry. This ensures that only noun phrases are considered keywords, and contextual keywords are given priority. As AKE methods are incapable of providing sufficient accuracy~\cite{papagiannopoulou2020}, we ask users to select the keywords relevant to the article out of ten identified keywords through the pop-up window shown in the Web browser, as depicted in Figure~\ref{fig:keywords}. Users are also allowed to add and select their own keywords through the user interface.

\subsection{Expert Opinion Discovery}

This step aims to explore the scientific views or comments of domain experts in the news media on the identified topic. The system combines the keywords selected by the user in the previous step with the \emph{AND} operator to generate a search query to search for relevant news items. Although a more useful query can be formed with a combination of different logical operators, we simply used the \emph{AND} operator for the sake of simplicity. The searches are done via Google's search APIs by considering the following types of news sources:

\begin{enumerate}
\item \emph{Mainstream News Outlets:} This includes credible news outlets with high traffic and wide news coverage. We set up a Google site-restricted search engine, which allows 10 websites for inclusion, and covered 10 news outlets with high credibility in English-language and having no paywall. For the source credibility measure, we considered the Media Bias/Fact Check (MBFC) credibility ratings\footnote{\url{https://mediabiasfactcheck.com/}} since it has been utilised by several recent studies~\cite{krieg2020, chen2020, weld2021}. The included news outlets are shown in Table~\ref{tab:newsoutlets}.

\item \emph{Scientific News Outlets:} This type involves credible pro-science news websites, featuring scientific views and recent research findings. For this part, we set up another Google site-restricted search engine with 10 selected news websites. The selection was made according to the MBFC credibility ratings with the help of bias, credibility, and traffic filters, and the websites with pro-science bias, wide news coverage, higher traffic, and no paywall were preferred. Table~\ref{tab:newsoutlets} indicates the list of selected websites of this type.

\item \emph{Other Credible News Sources:} In addition to the previous types, there are other types of news sources that might include expert opinions, such as news released by institutions and domain-specific news websites (e.g., Medscape, News Medical). To cover these, we set up a Google custom search engine without any site restriction to augment the search results containing the other two types of news sources. Since Google can also show results from non-news websites, we limited the search results with the \emph{NewsArticle}\footnote{\url{https://schema.org/NewsArticle}} Schema.org type to include only news articles. Furthermore, we utilised the Iffy Index of Unreliable Sources\footnote{\url{https://iffy.news/index/}}, which is based on MBFC, to exclude untrustworthy news sources from the search results.
\end{enumerate}

\begin{table*}[!htb]
\caption{News outlets covered by the site-restricted search engines}
\begin{tabularx}{\linewidth}{r X}
\toprule
\textbf{Mainstream News Outlets} & \textbf{Scientific News Outlets}\\
\midrule
NPR (\url{www.npr.org}) & Science (\url{www.science.org})\\
NBC News (\url{www.nbcnews.com}) & EurekAlert (\url{www.eurekalert.org})\\
Sky News (\url{news.sky.com}) & The Scientist (\url{www.the-scientist.com})\\
ABC News (\url{www.abcnews.go.com}) & Science News (\url{www.sciencenews.org})\\
Euronews (\url{www.euronews.com}) & MIT Technology Review (\url{www.technologyreview.com})\\
Reuters (\url{www.reuters.com}) & Popular Science (\url{www.popsci.com})\\
BBC News (\url{www.bbc.com}) & Science Daily (\url{www.sciencedaily.com})\\
PBS NewsHour (\url{www.pbs.com/newshour}) & Science Alert (\url{www.sciencealert.com})\\
Associated Press (\url{www.apnews.com}) & Live Science (\url{www.livescience.com})\\
CBS News (\url{www.cbsnews.com}) & The Conversation (\url{www.theconversation.com})\\
\bottomrule
\end{tabularx}
\label{tab:newsoutlets}
\end{table*}

The search results obtained from the three search engines are aggregated with the given order. Although it is possible to merge the three search engines into a single custom search engine, we preferred to use site-restricted engines for the first two types of news sources since we observed that site-restricted search engines provide more reliable results, and custom search engines configured to search the entire Web are limited to a subset of the Google Web Search corpus\footnote{\url{https://support.google.com/programmable-search/answer/70392}}. Hence, we benefited from a custom search engine as a secondary source to populate the obtained results from the site-restricted search engines.

\begin{figure*}[tb]
\centering
\includegraphics[width=0.8\linewidth]{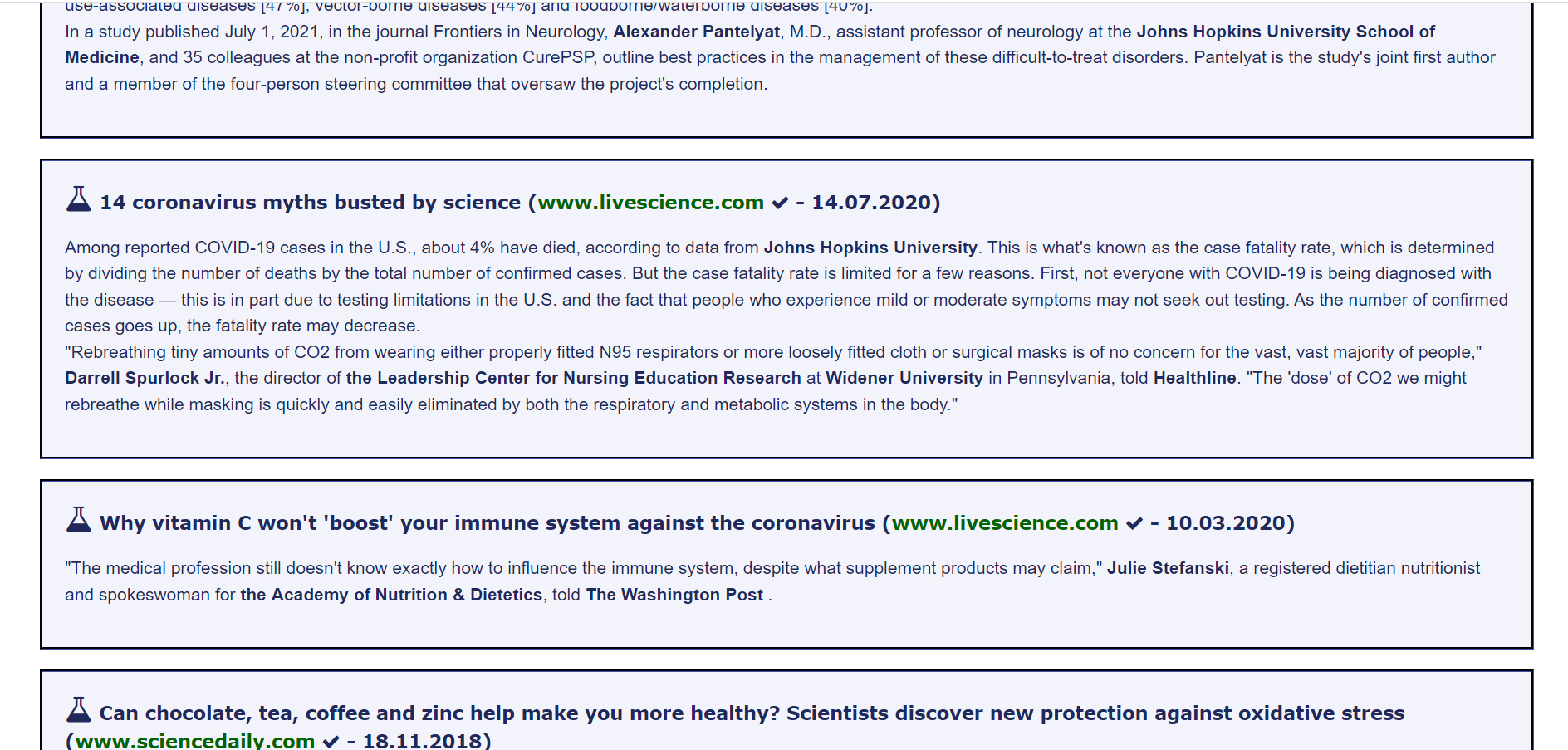}
\caption{An example output from aedFaCT showing some of the retrieved news articles.}
\label{fig:news}
\end{figure*}

Once the aggregated set of search results is obtained, the system tries to capture expert opinions from each article, which are mostly in the form of reported speeches, since they contain the most indicative elements (e.g., reported speeches, named entities, and quotes) of page usefulness for fact-checking~\cite{hasanain2022}. In this manner, it first downloads the news article with the Newspaper3k library. Then, the article is tokenised with two consecutive newline characters to obtain its paragraphs. Finally, for each paragraph, named entities are extracted with the spaCy library's NER feature. Only the paragraphs which contain at least one person name, one academic organisation name (containing an indicative word or phrase, such as \emph{university}, \emph{institute}, \emph{academy}, and \emph{research centre}) and a pair of single or double quotation marks (indicating a reported speech) are selected. As an exception, the summary extracted by the Newspaper3k library is shown to users for the \emph{The Conversation} news articles instead of retrieved expert opinions as they are already written by researchers and academics. As shown in Figure~\ref{fig:news}, the selected paragraphs are combined and shown to users in an individual box that also contains the source type (icon on the top-left), source name, and publish date. If the shown expert opinions are insufficient for a judgement and require further reading, users can click on the box to see the full article. Furthermore, a green clickable tick directing to the corresponding MBFC credibility rating webpage is added next to the names of the mainstream and science news sources for better explainability.

\subsection{Scientific Evidence Retrieval}

Scientific publications can also be considered a source of expert opinions as they are written by domain experts. Therefore, this step aims to retrieve research papers relevant to the topic of the input article.

\begin{figure*}[tb]
\centering
\includegraphics[width=0.8\linewidth]{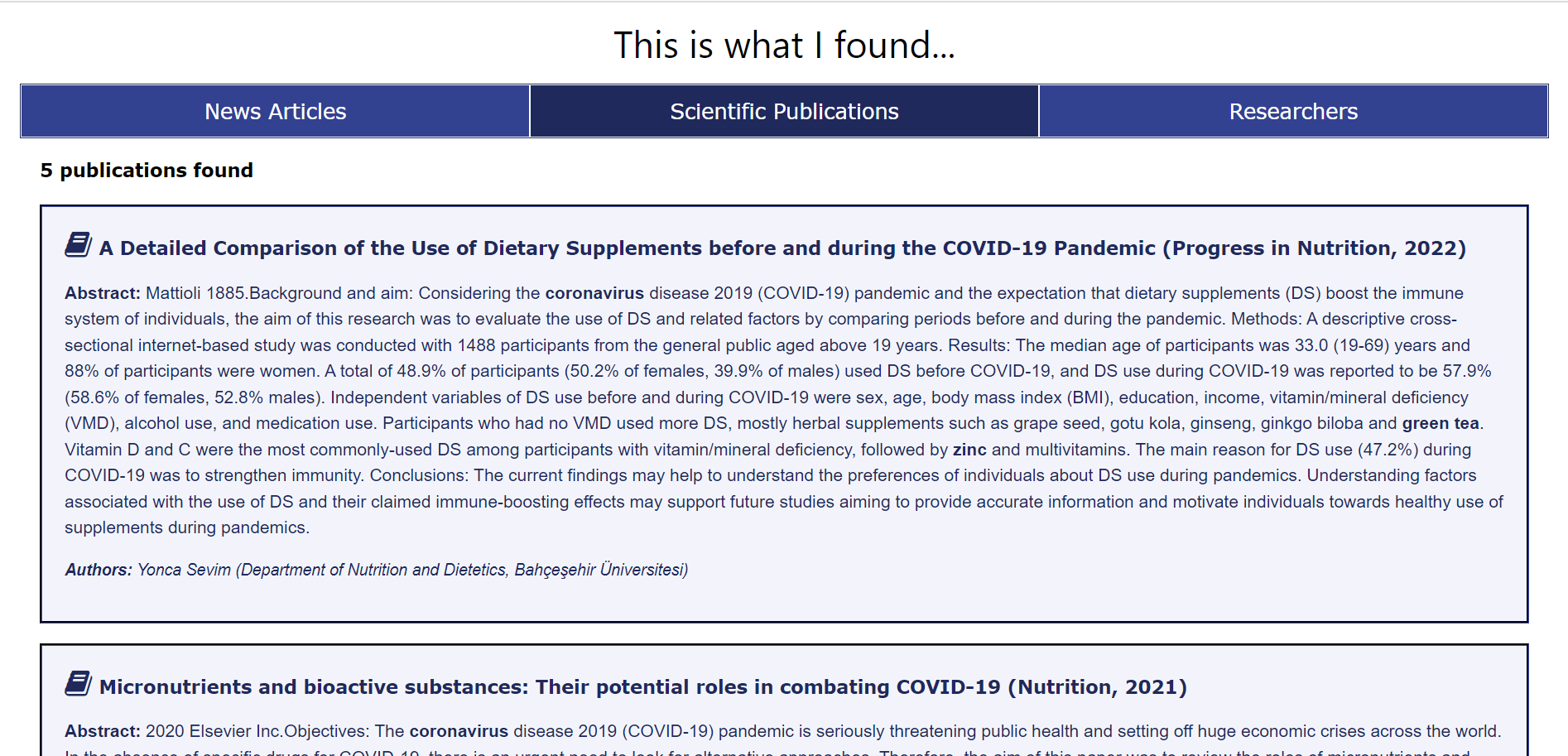}
\includegraphics[width=0.8\linewidth]{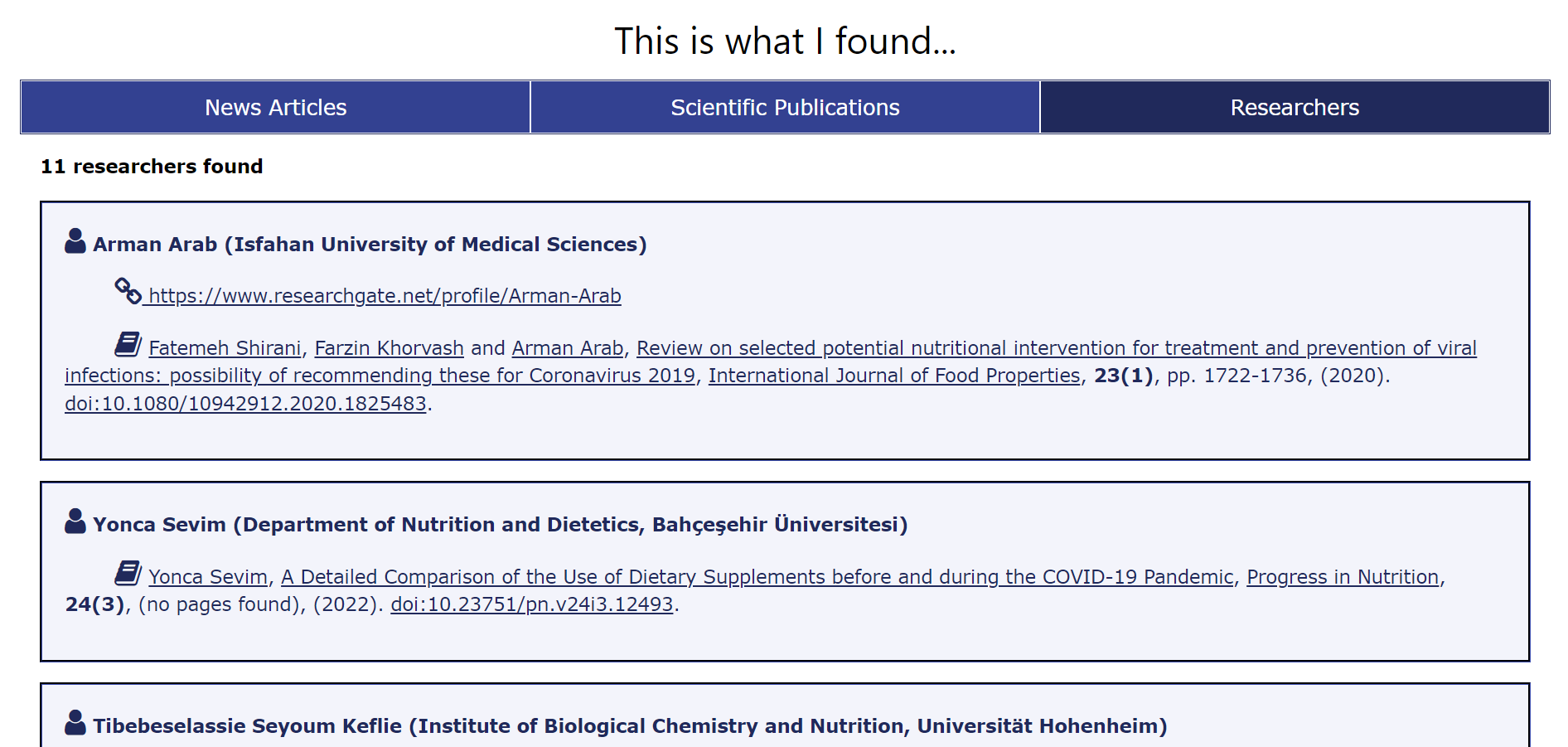}
\caption{An example output from aedFaCT showing some of the retrieved scientific publications and their co-authors, respectively}
\label{fig:researchers}
\end{figure*}

Similar to the previous step, we try to include only the records with high credibility. With this respect, we utilised Scopus API (with the help of the Pybliometrics library~\cite{Rose2019}) to search for relevant peer-reviewed publications. The searches are made by combining the selected keywords with an \emph{AND} operator, similar to the previous step. In addition, each keyword is surrounded by double quotation marks since it enables the inclusion of loose matches by allowing for wildcards and lemmatisation~\cite{scopus2022search}. As shown in the upper side of Figure~\ref{fig:researchers}, the obtained search results are shown to the user inside individual boxes containing the title, source, publication year, and abstract, with an order of relevance and publication year.

In addition to the scientific evidence provided by the tool, users, especially fact-checkers and journalists, might want to know the experts on the topic themselves to follow their research and/or make contact with them. To enable this, our proposed tool profiles the co-authors of the publications retrieved in the previous step, by obtaining relevant information, e.g., profile links, from their Scopus and ORCID profiles. The obtained researcher profiles are ordered by their number of publications in the search result. In the case that this number is equal, they are ranked based on the amount of information their profile contains to prioritise more contactable researchers. The bottom side of Figure~\ref{fig:researchers} shows an example output from the user interface showing a list of researchers.

\section{Evaluation}
\label{sec:evaluation}

To check the functionality and validity of the proposed Web browser extension, we conducted an initial evaluation as a pilot study with three co-authors (the third, fourth, and fifth) of this paper (one male and two female researchers), who were not included in the design and implementation phases of the tool. They were provided with 20 health news articles released by multiple sources with different credibility levels. The health domain was selected in order for a better simulation of common readers since it was outside the participants' areas of expertise. Then, the participants were asked to investigate the veracity of each news article and provide ratings for the shown output, in two rounds: 1) manually by following the investigation practices in their daily lives, such as using a Web search engine, and/or using a research database; 2) by using our proposed tool, aedFaCT.

For collecting the ratings from the participants, we set up a survey on Google Forms, containing a rating scale for each processed news article in both rounds together with a figure explaining each option in the scale. In addition, the survey included two questions to assess the perceived success of aedFaCT in terms of which approach had been faster and more helpful (with the options \emph{manual investigation}, \emph{investigation with aedFaCT}, and \emph{no difference}). Finally, it concluded with an open-ended question for comments and feedback.

In terms of the evaluation criteria in the rating scale, we followed Google's search quality guidelines~\cite{google2022}, which was proposed for evaluating Google search engine results with human raters. Although there are criticisms regarding the inadequacy of such retrieval effectiveness tests~\cite{lewandowski2015}, similar approaches are still being used in the literature~\cite{ciccone2015}. The guidelines involve mainly two tasks: determining to what extent the page achieves its purpose (``\emph{Page Quality}'') and determining if search results are useful (``\emph{Needs Met}''). Because our tool only benefits from credible news outlets and peer-reviewed publications, the former task is redundant in our case. Therefore, we only covered the latter task in our evaluation.

The ``Needs Met'' task involves two steps, which are about determining the user intent and the rating. Since all users of our tool will have the same intent, i.e., veracity assessment, the first step is redundant. Therefore, we only asked our evaluators to determine the rating of the search results by following the scale shown in Table~\ref{tab:ratingscale}.

\begin{table*}[!htb]
\caption{Rating scale for the Needs Met task~\cite{google2022}}
\begin{tabularx}{\linewidth}{r X}
\toprule
\textbf{Rating} & \textbf{Description}\\
\midrule
Fully Meets (\emph{FullyM}) & All or almost all users would be immediately and fully satisfied by the result and would not need to view other results to satisfy their need.\\
Highly Meets (\emph{HM}) & Very helpful for many or most users. Some users may wish to see additional results.\\
Moderately Meets (\emph{MM}) & Helpful for many users OR very helpful for some users. Some or many users may wish to see additional results.\\
Slightly Meets (\emph{SM}) & Helpful for fewer users. There is a connection between the query and the result, but not a strong\\
& or satisfying connection. Many or most users would wish to see additional results.\\
Fails to Meet (\emph{FailsM}) & Completely fails to meet the needs of the users. All or almost all users would wish to see additional results.\\
Not Applicable (\emph{N/A}) & The evaluator was unable to evaluate the result.\\
\bottomrule
\end{tabularx}
\label{tab:ratingscale}
\end{table*}

As a result of the evaluation, the average rating of the three raters when they manually investigated the given news articles was 4.35. This average has risen to 4.57 when they utilised aedFaCT in their investigations. In addition, the raters were in moderate agreement that aedFaCT provided better or similar results with respect to what they were able to obtain with their manual investigations, with a Fleiss' Kappa of 53.33\%. However, the raters have all agreed that fact-checking with aedFaCT was faster than their own practices. These results indicate that aedFaCT can help users perform fact-checking faster without degrading the quality of retrieved evidence for fact-checking. However, more extensive experiments are needed to evaluate its performance.

\section{Further Discussions}
\label{sec:discussions}

\subsection{Comparing aedFaCT with Existing Tools}

aedFaCT differs from existing fact-checking systems in several ways. To begin with, to the best of our knowledge, it is the first fact-checking system completely based on expert opinion discovery although there exist studies leveraging experts in fact-checking~\cite{labarbera2022, logically2022hamlet, wang2021}. Secondly, it is an evidence retrieval tool, and the final decision on the veracity is given by the user. Thus, it can establish trust among the users more easily unlike many fact-checking tools with a black-box design and fully automated decision-making mechanism, due to scepticism towards automation~\citep{juneja2022}. Another strength of aedFaCT is that it targets both common readers and professionals by retrieving both news articles and scientific publications. This enables users to consider information sources that they are more familiar with, depending on their level of expertise. In addition, it provides users with evidence from multiple sources and experts, which is a beneficial approach to breaking users out of their echo chambers. Finally, its overall workflow aligns with the common practices of human fact-checkers, in which engaging with experts is a key element, meaning that fact-checkers can use the tool to accelerate their claim investigation processes~\citep{juneja2022, micallef2022}. To be more precise, an overview of different characteristics of aedFaCT and other existing tools is provided in Table~\ref{tab:comparison}. As a result, we believe that aedFaCT can be made a new useful tool for fighting against false information and has the potential to be a part of standard fact-checking processes performed by both human fact-checkers and common readers.

\begin{table*}[tb]
\caption{The comparison between aedFaCT and other existing web browser extensions for fact-checking}
\begin{tabularx}{\linewidth}{l c c c c}
\toprule
\textbf{Tool} & \textbf{Approach} & \textbf{Task} & \textbf{Output} & \textbf{Domain}\\
\midrule
BRENDA & Automatic & Veracity Prediction & News articles & Any \\
FADE & Automatic & Veracity Prediction & News articles & Any \\
The Factual & Automatic & Credibility Assessment & Source \& content credibility & Any \\
InVID & Semi-automatic & Content Analysis & Additional information about content & Any \\
News2PubMed & Automatic & Evidence Retrieval & Research papers & Health only\\
NewsGuard & Automatic & Credibility Assessment & Source credibility & Any \\
News Scan & Semi-automatic & Credibility Assessment & Source \& content credibility & Any\\
\midrule
aedFaCT & Semi-automatic & Evidence Retrieval & News articles, research papers \& researchers & Any\\
\bottomrule
\end{tabularx}
\label{tab:comparison}
\end{table*}

\subsection{Limitations and Future Work}

The existing version of the proposed tool has a number of limitations. Firstly, it depends on external APIs (i.e., Google and Scopus) having quotas for the number of requests. Google Custom Search API\footnote{\url{https://developers.google.com/custom-search/v1/overview#pricing}} allows 10,000 requests per day while Scopus Search APIs\footnote{\url{https://dev.elsevier.com/api_ key_settings.html}} have a weekly quota between 5,000 and 20,000 requests, depending on the used API service. This makes it quite difficult to deploy aedFaCT for a wider community. Another limitation of the tool is its relatively low speed during keyword extraction. Since AKE methods already suffer from poor accuracy~\cite{papagiannopoulou2020}, we preferred accuracy over speed when selecting the AKE method and used the one (i.e., SIFRank+) providing the best accuracy although there exist various lightweight AKE algorithms. Besides, the information retrieval process was based on the selected keywords combined simply with the \emph{AND} operator. This causes fewer records as a result of the searches, especially when too many keywords were chosen by the user. Therefore, a smarter approach utilising a combination of logical operators (e.g., using the \emph{OR} operator for similar keywords) is needed for obtaining better search results. Lastly, the evaluation of aedFaCT has been conducted as a pilot study with a small number of participants having similar backgrounds. Hence, more extensive experiments with a more diverse and representative participant population, covering both professionals (e.g., fact-checkers and journalists) and common readers, are required.

Apart from resolving the limitations, for future work, we aim to improve the capabilities of aedFaCT. The existing version does not specifically consider retrieving results from official websites, e.g., governmental organisations, NGOs, and academic institutions. These can be retrieved by checking the URL extensions of the general Web results. Moreover, we plan to incorporate research on claim detection into aedFaCT so that extracted keywords can be more focused on specific claims in the input article.

\subsection{Broader Impact}

This work has some potential outcomes from a broader perspective. Since it accelerates the fact-checking process for users, it might encourage them to make fact-checking a daily activity and increase awareness in society for tackling false information online. However, users should be conscious when assessing the veracity of news items with the expert opinions shown by aedFaCT. Although aedFaCT retrieves evidence only from trustworthy sources, the displayed expert opinions might contradict each other due to disagreements between different experts. Therefore, aedFaCT does not eliminate the need for critical thinking ability for its users.

\section{Research Ethics Considerations}

The work reported in this paper involved a focus group discussion participated and a validation experiment participated by some co-authors of the paper only. According to the research ethics guidelines of the University of Kent's Central Research Ethics Advisory Group and general advice given by the School of Computing's Research Ethics Officer, such user studies involving researchers who are part of the research only were exempted from going through a research ethics review process. Both user studies did not involve any explicit collection of personal data or other sensitive data, and all participants explicitly consented to participate. Participating in the studies did not cause any noticeable harm to participants, but brought some benefits to them -- they could all achieve a better understanding of how to conduct fact-checking as a common reader and researcher.

\section{Conclusion}
\label{sec:conclusion}

Fake news is a challenging problem in society and causes serious harm. Its speed of propagation with the help of digital technologies suggests the need for automated solutions that can help people combat fake news at scale. Although there has been much effort to detect fake news, existing tools and services overlooked engaging with experts, who are commonly consulted during standard fact-checking processes. Therefore, this paper proposed \emph{aedFaCT}, a Web browser extension that retrieves expert opinions related to a news article to help fact-checkers and the general public perform fact-checking. Our initial evaluation suggested that it can accelerate fact-checking process without negatively affecting the search quality.

\subsubsection*{CRediT authorship contribution statement}

\textbf{Enes Altuncu}: Conceptualization, Methodology, Software, Writing -- original draft, Writing -- review \& editing. \textbf{Jason Nurse}: Methodology, Writing -- review \& editing, Supervision. \textbf{Meryem Bagriacik}: Investigation, Validation, Writing -- review \& editing. \textbf{Sophie Kaleba} :Investigation, Validation, Writing -- review \& editing. \textbf{Haiyue Yuan}: Validation, Writing -- review \& editing. \textbf{Lisa Bonheme}: Investigation, Writing -- review \& editing. \textbf{Shujun Li}: Conceptualization, Methodology, Supervision, Writing -- review \& editing.

\subsubsection*{Acknowledgements}

We would like to thank all the reviewers for their valuable feedback. The first and third co-authors, E.~Altuncu and M.~Bagriacik, were supported by funding from the Ministry of National Education, Republic of Turkey, through the MoNE-YLSY scholarship program.

\bibliography{main}

\end{document}